# The Value of Solving Pains

Jürg Meierhofer, Nikola Pascher, Jochen Wulf

Zurich University of Applied Sciences, CH 8401 Winterthur, Switzerland,
juerg.meierhofer@zhaw.ch, nikola.pascher@zhaw.ch, jochen.wulfj@zhaw.ch

Abstract. We introduce a novel framework aimed at identifying and quantifying the value of customer pains as a critical element in service innovation. The proposed approach enhances existing end-to-end frameworks by offering a structured method to elaborate on and measure the value derived from solving these customer challenges. The effectiveness of the framework is validated by operationalizing it in an industrial case study, where the model parameters were captured specifically and the value of solving various operational and structural pains was evaluated numerically.

Keywords: smart services · service innovation · value creation · service paradox.

## 1    Introduction

Buzzwords like the Industrial Internet of Things (IIoT) and Industry 4.0 highlight increasing opportunities for industrial companies using smart, data-based services. These services have significant potential for both designing new, value creating offerings and for solving customer pains, which can lead to substantially increased mutual value creation.

Generally, the economic benefit for companies using smart, data-based services are considered substantial. Not adopting these technologies might even risk losing competitive advantage. Despite being on the strategic agenda of nearly every major industrial company, implementation is slow, particularly in conservative industries where successful smart service implementations are few.

Decision makers are hesitant to invest upfront due to unclear value capture. Firms need to decide on heavy investments in digital infrastructure for new services. However, as the customer response to the new services typically unfolds in iterative development cycles over time, the return on these investments, i.e., value capture for the providers, is typically known ex-post, when the investment must already be considered sunk cost. This chicken-and-egg situation is described in (Meierhofer, Benedech, Schweiger, Barbieri, & Rapaccini, 2022). Additionally, the iterative process to develop new smart, data-based services and solve pains is lengthy. Both economic and technical aspects must be addressed crossfunctionally, which requires significant resources and slows decision-making.

The scientific literature shows a variety of methods for the development of services like (Bullinger, Ganz, & Neuhüttler, 2017; Schuh, Gudergan, & Kampker, 2016; Osterwalder, Pigneur, Bernarda, & Smith, 2014; Stickdorn, Hormess, Lawrence, & Schneider, 2018). There are also specific references focusing on the development of value creation by data-driven (smart) services (Schüritz & Satzger, 2016; Moser & Faulhaber, 2020; Jussen, Kuntz, Senderek, & Moser, 2019; Fruhwirth, Breitfuss, & Pammer-Schindler, 2020). However, there is a lack of methods to quantify the value of services in a measurable way. This can lead to the service paradox, i.e., the expected economic benefit not materializing after the service development (Gebauer, Fleisch, & Friedli, 2005).

(Meierhofer & Herrmann, 2018) describe an end-to-end approach for the development of data-driven services and qualitatively discuss how data-driven insights lead to a faster convergence of the innovation funnel (see Fig. 1). By quantitatively assessing the value of customer pains to be solved by a new service, the potential value creation by a service solving these pains is known. This will provide a basis for assessing the customers' willingness to pay for these services and thus the value that can be captured by the provider.

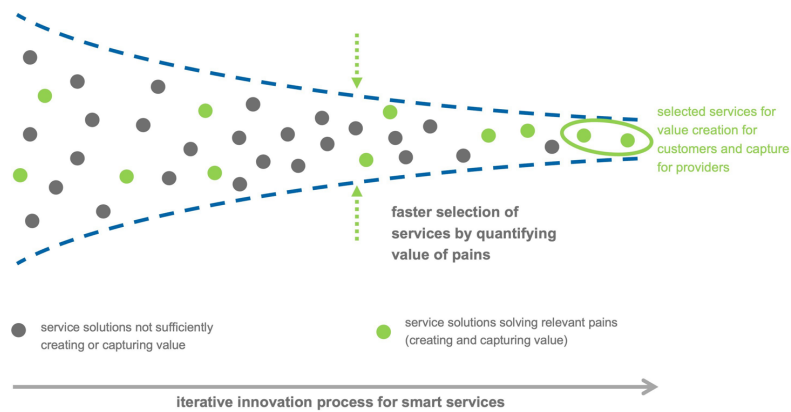

Fig.1. Faster convergence in the innovation funnel by quantifying value of pains (based on (Meierhofer & Herrmann, 2018))

Our framework discussed in this paper describes a modelling approach to identify the value of solving customer pains as a key step in service innovation. It provides a new approach for elaborating and quantifying this value and integrates this approach in the existing end-to-end service innovation frameworks. The validation of the new approach at a single case study with several use cases provides promising results and opens a multitude of perspectives for the further development of the methodology.

## 2    Prior Researcch

### 2.1     Value Quantification Practices for Value-based Pricing

Value quantification has been acknowledged by several authors as a prerequisite for value-based pricing. (Raja, Frandsen, Kowalkowski, & Jarmatz, 2020), for example, show that customer value analysis represents an important capability to enable value-based pricing and selling. Customer value analysis covers the analysis of customer processes and customer equipment data to calculate cost savings.

Firms traditionally have a cost-bias when evaluating product or service offers and calculate the total cost of ownership (TCO) including purchase cost and operating expenses (Snelgrove, 2017). However, when calculating TCO, there is an uncovered iceberg of cost drivers that often are neglected. The so-called hidden cost of ownership include maintenance cost, plant downtime, scrap, energy, and repair cost among others. (Snelgrove, 2017) suggests to take a lifecycle approach to assess the true TCO of a product or service (he coins the term "priceberg") and to include comparative levers that compare an offer to the status quo such as increased revenue because of higher end-product quality.

(Gray, Helper, & Osborn, 2020) introduce total value contribution (TVC) as a strategic sourcing approach that emphasizes maximizing an organization's long-term value rather than merely minimizing TCO. TVC measures various aspects of a product or service offering beyond just cost. Customer value includes factors such as product quality, innovation, and customization that enhance customer satisfaction. Revenue potential covers the potential for increased sales and market share through improved products or services. Furthermore, TVC includes risk related considerations of supply chain stability, disruption risks, and compliance with regulations and sustainability (environmental impact and social responsibility, such as reducing carbon footprint and ensuring ethical labor practices).

Regarding concrete quantification methods, literature distinguished ex-ante and ex-post approaches (Schuh, Leiting, Schrank, & Frank, 2022). Ex-post approaches measure the achieved value qualitatively with pilot customers or by continuously tracking benefits for data-driven services by analyzing customer usage data.

Ex-ante approaches include internal engineering, field value-in-use, indirect survey, focus groups, direct survey, benchmarks, conjoint analyses, and importance ratings (Anderson, Jain, & Chintagunta, 1992). Conjoint analysis is a tool that can be used to estimate customer value for a wide range of goods and services (Hinterhuber, 2004). This method involves presenting customers with two similar products that differ in price and other features and asking them to indicate their preferred attributes. This allows for the quantification of the value of specific product and service attributes for a group of customers. However, it requires a substantial customer response level to produce valid results, which is not feasible in many B2B customer scenarios.

Regarding high-priced industrial equipment, expert sales personnel may quantify aspects such as reduced failure rates, start-up costs, and life cycle costs, thereby demonstrating the value of a product to customers (Hinterhuber, 2004). (Jonas, Watkowski, Link, & Buck, 2023), for example, use a value lever framework to structure the value potential of a service across the entire value chain. Value levers are individual value components (e.g., reduction of unplanned maintenance work) of the value potential. The resulting value levers are then further specified and quantified, leading to an initial assessment of the monetization of the overarching digital services. As a second example, (Saccani, Alghisi, & Borgman, 2013) use expert judgements to model the expected monetary value of vessel remote monitoring. From the research discussed above it remains somewhat unclear how to derive components of customer value in a structured manner. Moreover, further methods are required that enable the quantification of the value components.

## 2.2     Methods for Quantifying Smart Service Value

Providers create value for customers by service provision. Inline with the concepts of value proposition and service design (e.g., (Osterwalder et al., 2014; Stickdorn et al., 2018), the design of such services starts with analyzing and understanding relevant customer pains. As we consider service provision in a B2B context, we focus on functional or financial pains of actors in the customer's organization - aware of the importance of other value components such as emotional or social. The pains may be operational such as a worker needing too much time to complete a task, missing relevant information or tools, operating a machine that does not perform su"ciently or break down too much, etc. Or the pains may be structural, such as a missing digital process for invoicing recurring revenues from services or other missing IT tools (see Fig. 3, pain 4).

The literature provides a multitude of schemes describing different dimensions for customer value creation (Leroi-Werelds, 2019). The functional and financial value dimensions considered in this paper are based on (Sweeney & Soutar, 2001), which also incorporates emotional and social value. The service value quantification of the latter two value dimensions is subject to future research and not considered in this paper. (Moody & Walsh, 1999) provide models for the valuation of information. (Moody & Walsh, 1999) as well as (Breuer et al., 2018; Möller, Otto, & Zechmann, 2017) differentiate between cost-based, marketbased, and utility-based valuation models. In our modelling context, information refers to the potential benefit of solving an operational pain and therefore, the utility-based perspective can be assumed. The information about the value of a pain enables the provider to offer a service that alleviates this pain. This is well aligned with the concept of functional and financial value according to (Sweeney & Soutar, 2001). (Meierhofer & Heitz, 2023) describe quantification methods for data-driven services that are based on these concepts.

## 3    Methodology

In a conceptual modelling step, we focus on relevant decision steps of the service innovation process and identify the quantification of the expected economic value to be created by new services as a relevant step. First, customer pains in a business-to-business (B2B) context are described in a formalized way that lends itself to a quantification in economic terms. Second, the value creation potential by solving these pains is analytically described. Based on this, the impact on value capture by the provider is integrated in the model.

These concepts are aligned with those from (Meierhofer & Heitz, 2023) and described in a consistent model. The effectiveness and practical applicability of the resulting quantitative model for mutual value creation based on solving pains are validated in a constructed single case study derived from real cases. This operationalization is intended to show how the model parameters are established in specific cases and how the value added is assessed numerically.

## 4    Model for Quantifying the Value of Pains

The model described in this paper formalizes mutual value creation in a B2B (business to business) provider-customer situation (Fig. 2) by solving pains. By informing the service design process about the functional and financial impact of these pains, the value created by solving these pains fully or partially by a smart service can be assessed. Therefore, in the sequel, the value of this information is formalized.

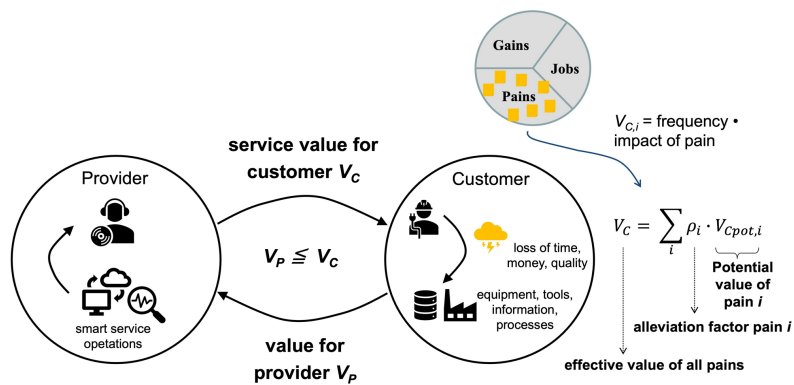

Fig.2. Conceptual model for value creation and capture by solving pains

This modelling approach is based on (Meierhofer & Heitz, 2023) and embeds this in the larger context of the service innovation process displayed in Fig. 1.

As indicated by Fig. 2, pains are formalized by their frequency of occurrence and by their impact. Considering a specific pain $i$, for instance a machine not performing properly because of a wear part not replaced sufficiently early, the negative value contribution of this pain for the customer is

$$V_{Cpot,i} = f_i \cdot v_i \tag{1}$$

where $f_i$ denotes the frequency of occurrence of the pain (e.g., once in 3 hours or weekly, monthly etc.) and $v_i$ the negative value created. For instance, if the reduced machine performance results in a capacity reduction for the hours until the wear part is replaced, this capacity reduction can be translated in a reduced number of pieces produced by the machine, which has a specific financial value. A smart service can therefore maximally create value in the amount $V_{Cpot,i}$ if it completely eliminates pain $i$. Thus, we denote $V_{Cpot,i}$ as the potential value creation by solving pain $i$. Fig. 3 shows a screenshot of a table as it is applied in practical projects for collecting the numerical values for the variables of Eq. 1.

In real situations, a service usually does not eliminate pains completely. E.g., some occurrences of pains may be missed out because the quality of data for detecting the pain (e.g., a worn part) is not sufficient. Therefore, we introduce the "alleviation factor" $\omega_i$ ($0 \rightarrow \omega_i \rightarrow 1$), with which the effective value creation for the customer becomes

$$V_{C,i} = \omega_i \cdot V_{Cpot,i} = \omega_i \cdot f_i \cdot v_i \tag{2}$$

with $V^{C,i} < V^{Cpot,i}$. It needs to be pointed out here that the pain alleviation factor $\omega_i$ incorporates a trade-off between cost and quality of the smart service. The effective value creation $V_{C,i}$ can asymptotically converge to the potential one $V_{Cpot,i}$ if the investment in data quality can be arbitrarily high. With higher investments, the probability of missing out on solving a pain can be reduced, i.e., the pain alleviation factor $\omega_i$ converges to 1. This means higher value creation, yet at higher costs, and makes the trade-off between cost and quality obvious. In the case with services that depend on data-based decision making, the pain alleviation factor can be directly related to the elements of the confusion matrix (true or false positives or negatives) (Provost & Fawcett, 2013). A potential smart service usually addresses several individual pains. Therefore, the total potential and effective value creation for the customer become:

$$V_{Cpot} = \sum V_{Cpot,i} \quad and \quad V_C = \sum V_{C,i} \tag{3}$$

Now, given that the value $V_C$ is created for the customer and this calculation can be made evident transparently, it becomes obvious that a rational business customer pays a service fee of at most $V_C$ to the provider. Thus, we have found an upper limit for the willingness to pay. In our model, value capture for the provider $V_P$ primarily comprises the sum of these payments, neglecting not directly financial components of value creation such as the value of customer data or customer loyalty. Therefore, we can state the relationship between value creation and value capture as:

$$V_P \leq V_C \qquad (4)$$

This simple relationship expresses that the provider cannot capture more value than it creates for the customer. The other way round, it also means that the provider can capture at most as much value is it creates, which may be higher or lower than its costs for providing the service. This opens perspectives on the economic value creation potential and on value-based pricing.

The total economic value creation by the service is the sum of the value creation for the customer and the provider, both in terms of potential and effective value creation:

$$V_{Economic,pot} = V_{Cpot} + V_{Ppot} \qquad and \qquad V_{Economic} = V_C + V_P \qquad (5)$$

The model described in this section enables the quantitative determination of value creation through smart services when they are applied to mitigate customer pains. On the one hand, this novel, quantitative approach expands the discussion of the value creation by smart services, which is usually conducted qualitatively in the literature. On the other hand, the approach also enables the selection of those service innovations that have a high value contribution as discussed in the Fig. 1. As a result, this model enables and supports the desired, effective and accelerated service innovation process. The trade-off between cost and quality with respect to service value creation will be discussed in section 6.

## 5    Application Example

Practical experience shows that there are numerous use-cases which can be evaluated in the postulated framework. Here we focus on firms which are active in producing goods with the help of machines in a typical IIoT or Industry 4.0 context. The following examples are based on real case studies, which are combined into a constructed case study for the purpose of anonymization and provided with modified numerical values. In the case study, sensor data from the producing machines is available, which can be used as a basis for smart service development. In Fig. 3 the examples 1 to 3 refer to operational pains, while example 4 denotes a structural pain. The pain examples refer to a manufacturing SME (small and medium enterprise) which produces specific components for its clients. From the perspective of providing additional service value (e.g., by an external service provider), we consider both the machine manufacturer and the operator of the machine as service beneficiaries, i.e., as customers.

With pain 1, the factory workers regularly (approximately once in two weeks) need to inquire a service desk because they lack information like some detail specification required for the completion of the production job. This creates a loss of one 1 hour (50 € assuming an hourly rate) for the factory worker and creates an additional work load of about 30 minutes (25 €) for the provider of the service desk. Applying eq. 2, this results in an annual effective value creation $V_{C,i=1}$ = 1500 € assuming that 80% of the problem cases can be avoided by the service. The same logic applies now to pains 2 and 3. With pain 2, about weekly the wear and tear of wearing parts is detected too late, which leads to a reduction in performance of approx. 1 hour of machine time (assuming 100 € machine costs per hour), with no impact on the provider side for this pain. With pain 3, about once in two months, the machine breaks down completely for 4 hours. This implies costs of 4 machine hours (400 €) and 4 worker hours (200 €) for the client, with total costs for support, logistics, and technician travelling of 1'000 € on the provider side. Together, a service provider can create value for the customer and the provider of annually 11'220 € in total by solving these operational pains with the alleviation factors indicated in Fig. 3. Given Eq. 4, this means that the service provider can charge a price of at most 11'220 € per year for this service (i.e., value-based pricing), with typical pricing targeting at 50 : 50 revenue sharing.

| Pain nr. | Pain description | Value Creation for Agent of Customer | | | | Value Creation for Agent of Provider | | |
|---|---|---|---|---|---|---|---|---|
| | | Frequency (annual) | impact € | Alleviation Factor | Value of pain (annual) € | Frequency (annual) | impact € | Value of pain (annual) € |
| **Total annual value creation by solving operational pains** | | | | | **6'520** | | | **4'700** |
| 1 | Missing information about current job => technical service desk inquiries | 25 (appr. once per 2 weeks) | 50 (1 hour search time) | 0.8 | 1'000 | 25 | 25 (30 minutes technical service agent time) | 0.8 | 500 |
| 2 | Low machine performance due to wear parts not being replaced timely | 50 (alomost weekly) | 100 (1 hour of peformance) | 0.6 | 3'000 | - | - | - | - |
| 3 | Machine break downs | 6 (once per 2 monts) | 600 (4 hours machine costs + idle operator) | 0.7 | 2'520 | 6 | 1'000 (technician, logistics, travelling) | 0.7 | 4'200 |
| **Total annual value creation by solving structural pain 4** | | | | | **1'260** | | | **600** |
| 4 | Recurring revenue can not be billed because of missing IT tool | 12 (assuming a monthly payment) | 150 (3 hours additional effort for workarounds) | 0.7 | 1'260 | 12 | 100 (2 hours additional effort for workarounds) | 0.5 | 600 |

Fig.3. Example listing of value of pains as assessed in practical application cases.

Structural pain 4 describes a situation in which there is a lack of processes and accounting systems to charge recurring usage fees to the customer. This is a situation regularly encountered in practice with SME's which are in the transformation from selling products to selling "product usage as a service". They don't have systems and processes in place for recurring billing, either on the side of the provider or the customer or both. Yet, they have workarounds for manually creating these recurring bills, however at the monthly costs of 3 hours (150 €) on the customer side and 2 hours (100 €) on the provider side. A service provider can now reduce this effort to zero by a service automating this billing process in 70% of the cases on the customer side (alleviation factor 0.7) and 50% of the cases (alleviation factor 0.5) on the provider side. This creates total annual value of 1'860 € and thus defines the maximum price that can be charged for this service.

## 6    Conclusions and Recommendations

The quantitative model to assess the economic value creation $V_{Economic}$ by a smart service idea as well as its impact on value-based pricing are applied in the service innovation process shown in Fig. 1. The model allows to assess whether a new smart service idea creates sufficient value including its costs and unfolds in the four steps shown in Fig. 4:

1. For a new service idea in the flow of the innovation process (circled in the example of Fig. 4), the economic value created and captured by solving customer pains is assessed as described in section 4. This results in the total economic value created $V_{Economic}$.

2. This value $V_{Economic}$ is evaluated against the costs for developing and operating the service. Generally, this evaluation yields that either i) the economic value created $V_{Economic}$ could or should be improved by an improved service design, which means higher costs (e.g., better data quality), or ii) the economic value created $V_{Economic}$ is good enough but the costs are too high, or iii) both the economic value created $V_{Economic}$ is not sufficient and costs are already too high.

3. While scenario iii) described above means that the new service idea is immediately abandoned, scenarios i) and ii) result in a redesign of the service. Either a more sophisticated design is researched, improving $V_{Economic}$, but also highering costs. Or a more e"cient and lean design is worked out, reducing costs while maintaining a su"cient pain relieving for keeping $V_{Economic}$ sufficiently high.

4. The adapted service design is re-entered in the innovation funnel and assessed against other service innovation ideas. If this assessment turns out that the redesigned service is still in-line with the innovation targets, it can be taken to the next innovation stage gate or, if a re-evaluation is required, it can again undergo the quantitative value assessment cycle by re-entering step 1 of this procedure.

Our use case example addresses only functional pains which can be resolved by a technical solution (a smart service) for the customer. The framework can be extended to estimate the value of a variety of other pains, which may be non-functional, such as:

– Environmental impacts, which might harm nature and / or lead to an increased CO2 emission
– Social recognition or a reputation risk to the customer which might lead to a decreased popularity of the customer's products and thus to decreased revenue

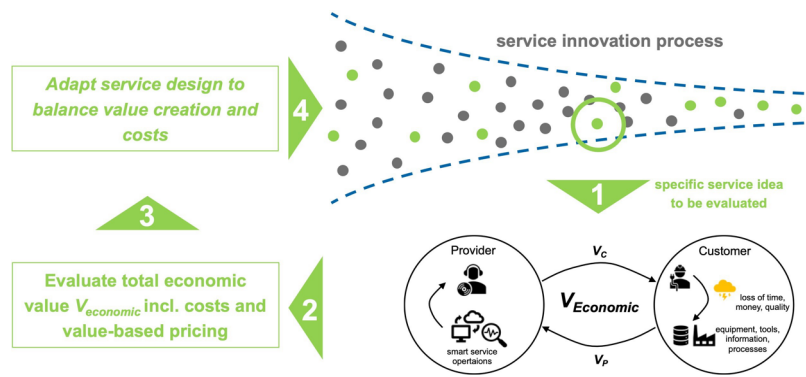

Fig.4. Quantitative assessment of economic value $V_{Economic}$ of service idea to achieve faster convergence shown in Fig. 1

- A threat to the company culture, which could lead to a loss of skilled workforce

The new service development framework extends the scientific literature of service innovation by a quantitative modelling approach. It thus opens new perspectives on value creation by services. Although the application example is drawn from an SME environment, the model can equally be applied to any organization type or use case. The model could have a high impact in a corporate context, where significant investment decisions in innovative digital technologies need to be taken at a point in time when only a rough idea of the value proposition is known ex-ante. Quantifying the value of a resolved customer pain and combining it with consecutive pricing schemes allows to directly translate it into the company's business potential, thus basing the investment decision on a quantitative basis.

## References

Anderson, J. C., Jain, D. C., & Chintagunta, P. K. (1992, June). Customer Value Assessment in Business Markets: A State-of-Practice Study. *Journal of Business-to-Business Marketing*, *1*(1), 3–29. Retrieved 202407-24, from http://www.tandfonline.com/doi/abs/10.1300/J033v01n01_02     doi: https://doi.org/10.1300/J033v01n01_02

Breuer, M., Greif, C., Klekamp, J., Lippert, S., Meier, S., Mies, S., ... Wegmann, T. (2018). *Data Economy - Datenwertschöpfung und Qualität von Daten.*

Bullinger, H., Ganz, W., & Neuhüttler, J. (2017). Smart Services – Chancen und Herausforderungen digitalisierter Dienstleistungssysteme für Unternehmen. In M. Bruhn & K. Hadwich (Eds.), *Dienstleistungen 4.0.* Wiesbaden: Springer.

Fruhwirth, M., Breitfuss, G., & Pammer-Schindler, V. (2020, January). The Data Product Canvas - A Visual Collaborative Tool for Designing DataDriven Business Models. BLED 2020 Proceedings. Retrieved from

https://aisel.aisnet.org/bled2020/8

Gebauer, H., Fleisch, E., & Friedli, T. (2005, February). Overcoming the Service Paradox in Manufacturing Companies. European Management Journal, 23(1), 14–26. doi: https://doi.org/10.1016/j.emj.2004.12.006

Gray, J. V., Helper, S., & Osborn, B. (2020). Value first, cost later: Total value contribution as a new approach to sourcing decisions. Journal of Operations Management, 66(6), 735–750. Retrieved 202408-08, from https://onlinelibrary.wiley.com/doi/abs/10.1002/joom.1113 (_eprint: https://onlinelibrary.wiley.com/doi/pdf/10.1002/joom.1113) doi: https://doi.org/10.1002/joom.1113

Hinterhuber, A. (2004, November). Towards value-based pricing—An integrative framework for decision making. Industrial Marketing Management, 33(8), 765–778. doi: https://doi.org/10.1016/j.indmarman.2003.10.006

Jonas, C., Watkowski, L., Link, J., & Buck, C. (2023, December). Identifizierung von Wertpotenzial für digitale Services von B2BFertigungsunternehmen am Beispiel eines Küchengeräteherstellers. HMD Praxis der Wirtschaftsinformatik, 60(6), 1328–1344. Retrieved 202407-23, from https://link.springer.com/10.1365/s40702-023-00964-2 doi: https://doi.org/10.1365/s40702-023-00964-2

Jussen, P., Kuntz, J., Senderek, R., & Moser, B. (2019, January). Smart Service Engineering. Procedia CIRP, 83, 384–388. Retrieved 2021-01-07, from http://www.sciencedirect.com/science/article/pii/S2212827119307036 doi: https://doi.org/10.1016/j.procir.2019.04.089

Leroi-Werelds, S. (2019, January). An update on customer value: state of the art, revised typology, and research agenda. Journal of Service Management, 30(5), 650–680. Retrieved 2021-01-06, from https://doi.org/10.1108/JOSM-03-2019-0074 (Publisher: Emerald Publishing Limited) doi: https://doi.org/10.1108/JOSM-03-2019-0074

Meierhofer, J., Benedech, R., Schweiger, L., Barbieri, C., & Rapaccini, M. (2022). Quantitative Modelling of the Value of Data for Manufacturing SMEs in Smart Service Provision. ITM Web of Conferences, International Conference on Exploring Service Science (IESS 2.2), 41, 04001. (Publisher: EDP Sciences) doi: https://doi.org/10.1051/itmconf/20224104001

Meierhofer, J., & Heitz, C. (2023, October). On the Value of Data in Industrial Services: How to Optimize Value Creation by Reconfiguration of Operant Resources. Journal of Creating Value. (Publisher: SAGE Publications India) doi: https://doi.org/10.1177/23949643231199002

Meierhofer, J., & Herrmann, A. (2018). End-to-End Methodological Approach for the Data-Driven Design of Customer-Centered Digital Services. In G. Satzger, L. Patrício, M. Zaki, N. Kühl, & P. Hottum (Eds.), Exploring Service Science (pp. 208–218). Cham: Springer International Publishing. doi: https://doi.org/10.1007/978-3-030-00713-3_16

Moody, D. L., & Walsh, P. (1999). Measuring the Value Of Information-An Asset Valuation Approach. In ECIS (pp. 496–512).

Moser, B., & Faulhaber, M. (2020). Smart Service Engineering. In M. Maleshkova, N. Kühl, & P. Jussen (Eds.), Smart Service Management:

Design Guidelines and Best Practices (pp. 45–61). Cham: Springer International Publishing. doi: https://doi.org/10.1007/978-3-030-58182-4_5

Möller, K., Otto, B., & Zechmann, A. (2017). Nutzungsbasierte Datenbewertung. Controlling, 29(5), 57–66. doi: https://doi.org/10.15358/0935-0381-20175-57

Osterwalder, A., Pigneur, Y., Bernarda, G., & Smith, A. (2014). Value Proposition Design: How to Create Products and Services Customers Want. John Wiley & Sons.

Provost, F., & Fawcett, T. (2013). Data science for business. Sebastopol, Calif: O'Reilly.

Raja, J. Z., Frandsen, T., Kowalkowski, C., & Jarmatz, M. (2020). Learning to discover value: Value-based pricing and selling capabilities for services and solutions. Journal of Business Research, 114, 142–159. (Publisher: Elsevier)

Saccani, N., Alghisi, A., & Borgman, J. (2013). The Value and Management Practices of Installed Base Information in Product-Service Systems. In C. Emmanouilidis, M. Taisch, & D. Kiritsis (Eds.), Advances in Production Management Systems. Competitive Manufacturing for Innovative Products and Services (pp. 415–421). Berlin, Heidelberg: Springer. doi: https://doi.org/10.1007/978-3-642-40361-3_53

Schuh, G., Gudergan, G., & Kampker, A. (Eds.). (2016). Management industrieller Dienstleistungen. Berlin, Heidelberg: Springer Berlin Heidelberg. doi: https://doi.org/10.1007/978-3-662-47256-9

Schuh, G., Leiting, T., Schrank, R., & Frank, J. (2022). Value-based Pricing von Smart Services im Maschinen- und Anlagenbau. In M. Bruhn & K. Hadwich (Eds.), Smart Services (pp. 255–275). Wiesbaden: Springer Fachmedien Wiesbaden. (Series Title: Forum Dienstleistungsmanagement) doi: https://doi.org/10.1007/978-3-658-37346-7_9

Schüritz, R., & Satzger, G. (2016, August). Patterns of Data-Infused Business Model Innovation. In 2016 IEEE 18th Conference on Business Informatics (CBI) (Vol. 01, pp. 133–142). (ISSN: 2378-1971) doi: https://doi.org/10.1109/CBI.2016.23

Snelgrove, T. (2017). Creating, calculating and communicating customer value: How companies can set premium prices that customers are willing and able to pay. In Innovation in Pricing (pp. 248–260). Routledge.

Stickdorn, M., Hormess, M. E., Lawrence, A., & Schneider, J. (2018). This Is Service Design Doing: Applying Service Design Thinking in the Real World. "O'Reilly Media, Inc.". (Google-Books-ID: aqRGDwAAQBAJ)

Sweeney, J. C., & Soutar, G. N. (2001, June). Consumer perceived value: The development of a multiple item scale. Journal of Retailing, 77(2), 203–220. doi: https://doi.org/10.1016/S0022-4359(01)00041-0